\documentclass[aps,pra,showpacs,amssymb,amsmath,nobibnotes,a4paper, 10pt,twocolumn]{revtex4-2}

\usepackage[dvips]{epsfig}
\usepackage{xcolor,soul}
\usepackage{graphics,graphicx}
\usepackage{epstopdf}
\usepackage{hyperref}
\usepackage{mhchem}
\usepackage{bm}
\usepackage{ulem}

\DeclareGraphicsExtensions{.jpg,.pdf,.mps,.png}
\renewcommand{\underline}{\ul}

\begin{document}

\title{Thermal bistability in laser-cooled trapped ions} 
\author{A. Poindron, J. Pedregosa-Gutierrez, C. Champenois}
\affiliation{Aix-Marseille Université, CNRS, PIIM, Marseille, France}
\date{\today}
\vspace{-12em}
\begin{abstract}
The non-linear dynamics of large ion clouds ($N \geq 256 $ ions) trapped in radio-frequency traps and coupled to laser-cooling give rise to a bistable behaviour of the temperature.  Numerical simulations of the free evolution of a large three-dimensional spherical cloud  is used to characterise  how the oscillating field amplitude and the ion number  control the rf heating rate and the switching between the two stable states. We show that the heating rate does not significantly depends on the ion number but strongly depends on the oscillating field. This exhibits the major role played by the density of the ensemble. The competition between the radio-frequency heating and the laser cooling is also discussed. They are used to design scenarii to take advantage of the threshold effect to detect and quantify perturbations by an intruder.

\end{abstract}

\maketitle 

\section{Introduction}
Ion traps are experimental tools used to manipulate atomic or molecular ions in quantities ranging from a single ion \cite{wineland87} to thousands of ions \cite{prestage91}.

Experiments based on large ion cloud remain a very relevant choice  for micro-wave frequency standards \cite{prestage91}, high precision spectroscopy of cold molecular ions \cite{willitsch08r,alighanbari18,alighanbari20}, chemical reaction in the cold and diluted limit \cite{yang21} or quantum optics in the large cooperative regime \cite{dantan09,dantan10,laupretre19}. In these experiments where the cold ion cloud expands in the three directions, the rf-driven motion (also known as micro-motion) is one of the factors  limiting  resolution and accuracy in spectroscopy because of the induced Doppler shift  \cite{prestage89,berkeland98,prestage06}. It is also the source of the radio-frequency heating, an energy transfer from the rf source to the thermal motion of the ions, mediated by the Coulomb interaction. This rf-heating is responsible for fast transitions from low to high temperature state, referred as an "explosive onset" in the first studies about chaotic trajectories of few ions system \cite{hoffnagle88}. This is the signature of a strong non-linear dynamics driven by the Coulomb interactions.  So far, few molecular dynamics simulations have studied such features like  in  \cite{prestage91rf}, where the dynamics of ion crystal as large as 512 in a 3D-Paul trap concludes that ion clouds at 5 mK show rapid heating when the rf electric field amplitude is large enough. This strong dependence with the amplitude of the rf electric field was confirmed in \cite{Ryjkov05} where periodic boundary conditions are used to study the translationally uniform ion cloud geometry that is a relevant representation for very elongated cloud in 2D linear quadrupole traps. The work detailed in \cite{tarnas13,nam14,nam17} goes beyond the microscopic description and proposes a universal heating formula to capture the dependence with the rf electric field, ion number and temperature for spherical cloud in 3D-Paul trap. 

Even if most experimental signals rely on the laser induced fluorescence involved in Doppler laser cooling, few works have combined the study of rf-heating in the context of laser-cooling \cite{marciante10,poindron21}. As shown in this article, when Doppler laser-cooling is added to this complex system, a bistable behaviour can be observed, where the control parameter is the strength of the rf trapping field and the switching parameter is the temperature of the sample. This bistability has a strong impact on experiments involving large samples like shown in the last part of this article.
Here, we  study this bistability by computing the rf-heating of finite size ion ensembles by means of molecular dynamics simulations, using the data generated during the free evolution of an ion cloud confined in a linear rf-trap.  Rf-heating rates  are then directly compared to laser cooling rates to demonstrate the onset of a bistable behaviour for the temperature of the sample. Investigating the effect of the ion number $N$ and the amplitude of the radio-frequency field we demonstrate that only the latter has a significant effect on rf heating rate,  pointing to the cloud density as the controlling parameter. These results are then used to build and justify experimental scenarii based on the control of the switch between the two stable states.

\section{Molecular dynamic simulations}\label{s:model}
\subsection{Model}
The ion ensemble is confined in a linear quadrupole rf trap (inner radius $r_0$=2.5~mm) where each diagonal pair of rods, aligned with the $Oz$ axis, is supplied with  time oscillating voltage (frequency $\Omega/2\pi = 2$ MHz) with opposite phase and a common amplitude $U_{RF}$ (for more details, see \cite{poindron21}).  The potential generated in the plane perpendicular to the trap axis is
\begin{equation}
\Phi(x,y,t) = \frac{U_{RF}\cos(\Omega t)}{r_0^2}\left( x^2 - y^2 \right). \label{eq:rad_pot}
\end{equation}
A static voltage $U_{DC}$ is applied to extra electrodes to generate a quadratic potential $\Phi(z)$ along the  $Oz$ axis, characterised by the frequency $\omega_z/2\pi$ scaling as $\omega_z/2\pi =100 \times \sqrt{U_{DC}}$~kHz for trapped Ca$^+$ ions, of mass $m=40$~a.m.u. and charge $Q = +q_e = + 1.602\cdot 10^{-19}$~C. 
The equations governing the ion dynamics read
\begin{align}
&\ddot{u}_i + \left( -\frac{\omega_z^2}{2} \pm \frac{2 Q U_{RF} \cos{\Omega t}}{m r_0^2} \right)u_i \notag \\
&\quad\quad\quad\quad = \frac{Q^2 }{4\pi \epsilon_0 m}\sum_{j=1,j\neq i}^N{\frac{u_i - u_{j}}{|\bm{r}_{i}-\bm{r}_{j}|^3}}  \label{eq:dif1}\\
&\ddot{z}_i + \omega_z^2 z_i = \frac{Q^2 }{4\pi \epsilon_0 m}\sum_{j=1,j\neq i}^N{\frac{z_i - z_{j}}{|\bm{r}_{i}-\bm{r}_{j}|^3}} \label{eq:dif}
\end{align} for an ion $i$, where $u_i$ stands for $x_i$ or $y_i$ and $\bm{r}_i = (x_i,y_i,z_i)$. Those that govern the motion in the radial plane $(Ox,Oy)$ can be recast into the standard Mathieu equations \cite{McLachlan47} by defining the usual $q_x$ Mathieu parameter giving the dimensionless rf-traping electric field and that we choose positive $q_x = (4QU_{RF})/(mr_0^2\Omega^2)$ and the effective Mathieu parameters $a_x=a_y$  defined as  $a_x = -2\omega_z^2/\Omega^2$ \cite{drewsen00,drakoudis06}.

{The equivalent static pseudo-potential \cite{major68} is defined by the fundamental oscillation frequencies  for a single ion in the radial direction $\omega_r$ and in the axial direction $\omega_z$. By changing the strength of the rf trapping field, characterised by the $q_x$ parameter, two characteristics of the ion clouds are changed. They are the mean density, which scales like $q_x^2$ \cite{prasad79, hornekaer02} and the geometrical aspect ratio, which depends on the potential aspect ratio $\omega_z^2/\omega_r^2$.   To have only one characteristic changing along the simulations, the   ratio $\omega_z^2/\omega_r^2$  is chosen equal to 1 for all the tested Mathieu parameter $q_x$, which implies a spherical shape to the ion cloud \cite{hornekaer02}. To achieve this goal, we compute the continuous fraction \cite{McLachlan47,major05} defining the stability parameter $\beta_x( a_x, q_x)= 2 \omega_r /\Omega$ recursively to impose $\beta_x^2(a_x, q_x)= -2 a_x$.

\subsection{Temperature}
As the rf-driven motion does not contribute to the temperature, it must  not be taken into account in its definition \cite{Schiffer00,marciante10}. To that purpose, the velocity ${\bm{v}_i}(t)$ of each ion $i$ is averaged over one radio-frequency period, to smear out the rf-driven oscillation. The temperature $T$ is dependant on this time-averaged velocities $\overline{\bm{v}_i}(t)$ like
\begin{equation}
\frac{3}{2}k_BT = \frac{1}{2}  \frac{m}{N}\sum_{i=1}^N \overline{\bm{v}_i}(t)^2  \label{eq:T}
\end{equation}
with $k_B$ the Boltzmann's constant. Fig.~\ref{fig:Tcurve} shows the time evolution for this temperature for a sample made of 1024 ions, following a thermalisation process that takes the sample to  10~mK \cite{poindron21}. After few milliseconds of free evolution, we observe a sudden increase of the temperature by two orders of magnitude. We have checked that the velocity distribution remains in a very good agreement with a Maxwell-Boltzmann distribution along this evolution (Fig. \ref{fig:Tcurve}).
\begin{figure}[htb]
    \includegraphics[width=0.5\textwidth]{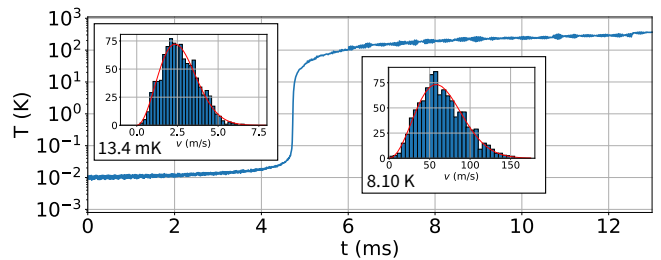}
  \caption{Temperature evolution of a 1024 ion ensemble, in a trap defined by $q_x=0.64$ and $\omega_z/2\pi = 388$~kHz.  The temperature is computed following Eq. (\ref{eq:T}).  The two insets show the velocity distribution of a 1024 ion ensemble at 13.4~mK and 8.10~K as histogram. The red curve is the computed Maxwell-Boltzmann distribution at this given temperature. 
  \label{fig:Tcurve}}
\end{figure}

\section{Radio-frequency heating and laser cooling}\label{s:heating}
\subsection{Heating rate}
To identify the general characteristics  of these sudden temperature rise, the heating rate $H=\mathrm{d}T/\mathrm{d}t$ is calculated based on a time averaged temperature,  to smooth the fluctuations induced by the small size of the sample. The averaging time is chosen at 100~$\mu$s which correspond to $n_t=200$ radio-frequency periods $\tau_{RF}$.

Fig.~\ref{fig:heating_rate} shows these heating rates depending on the temperature computed for a cloud of $N=1024$ ions, trapped by five different trapping field amplitude, characterised by Mathieu parameters $q_x=\{0.2, 0.3, 0.4, 0.5, 0.6\}$. Typically, a heating rate curve represented in log-log exhibits the following behaviour : at low temperature the heating rate log first increases linearly with the temperature log, going from below $10^{-6}$ K/$\tau_{RF}$ to the range of $10^{-2}$ K/$\tau_{RF}$ for a temperature in the range 0.01-0.1~K. This is the signature of a polynomial dependence of the heating rate with the temperature, with an  exponent increasing with $q_x$. Then the heating rate remains stationary within the range $10^{-2}-10^{-1}$ K/$\tau_{RF}$ up to $10$ K and then decreases smoothly down to 400 K where simulations are stopped.  

These curves are in accordance with the $\Lambda$-shape scheme proposed in \cite{Blumel89} to explain the equilibrium temperature of a trapped ion cloud. For $q_x=0.4$, we find heating rates as a function of temperature in accordance with the results of \cite{Ryjkov05}, computed for $q_x=0.44$. However, we could not confirm that heating rate behaves universally independently of the Mathieu parameter $q_x$, as it was suggested in Ref.~\cite{tarnas13}. 

\subsection{Laser-cooling}
To characterize the equilibrium thermal state of a laser-cooled sample, we now compare the rf-heating rate to the Doppler laser cooling rate $G$. Doppler laser-cooling allows to cool down ions thanks to the resonant radiation pressure induced by the recoil when ions absorb photons \cite{hansch75,Wineland78}. The laser cooling of Ca$^+$ ions involves two transitions and three levels but we consider here the simplified model of a two-level system, keeping only the resonant dipole transition at $\lambda_L = 397$~nm to the first excited state with lifetime $\tau_e=6.9$~ns \cite{hettrich15}. This simplification over-estimate the optimum probability $P_e$ for an ion to be in the excited state by a factor close to 2 \cite{lisowski05} but it will not change the conclusion of the comparison. For a laser beam with wave-vector $\bm{k}_L=2\pi/\lambda_L \bm{z}$, propagating along the symmetry axis $Oz$ of the trap, the scattering force is \cite{metcalfbook} $F_s =  \Gamma P_e \hbar {\bm k}_L $ with $\Gamma=1/\tau_e$. The probability $P_e$ depends on the Rabi frequency $\Omega_R$ of the laser-dipole coupling, relative to the natural spontaneous emission rate $\Gamma$, by the on-resonance saturation parameter $s=2\Omega_R^2/\Gamma^2$. It depends also on the laser frequency $\omega_L$ by its detuning $\delta$ from the atomic transition frequency $\omega_0$ as
\begin{equation}
P_e = \frac{1}{2}\frac{s}{1+s+4\delta^2/\Gamma^2}.
\end{equation}
For a moving atom, the detuning must include the Doppler effect and $\delta = \delta_L - \bm{k}_L\cdot \bm{v}$, with $\delta_L = \omega_L - \omega_0$. The work rate (or power) of the radiation pressure force on an ion with velocity $\bm{v}$ can be calculated  as $\bm{F_s} \cdot \bm{v}$,  assuming that the excitation probability $P_e$ has reached the stationary limit for a given velocity. This assumption is justified because the transition sets in the broadband limit where $\Gamma \gg k_L v_r$ where $v_r=\hbar k_L/m$ is the recoil velocity due to the absorption of one photon \cite{champenois16}. This work rate,  averaged over the ion velocity distribution $P(\bm{v},T)$ of the sample, defines the cooling rate that we choose positive and express in temperature variation per unit time:
 \begin{equation}
G = - \frac{1}{k_B}\int \left(\bm{F_s} \cdot \bm{v}\right) \  P(\bm{v},T) ~\mathrm{d}\bm{v}. 
\label{eq:G}
\end{equation}
For the calculation shown on Fig.~\ref{fig:heating_rate}, a Maxwell-Boltzmann distribution is assumed and the cooling rate is calculated for each temperature, for a saturation parameter $s=2$ and laser detunings $\delta_L= -\Gamma$ and $\delta_L=-10 \Gamma$.  To be complete, the heating and cooling rate comparison must include the heating rate $H_{e}$ induced by the spontaneous emission involved in the laser cooling. This can be calculated as \cite{lett89}
 \begin{equation}
H_e =  \frac{\hbar^2 k_L^2}{m k_B} \int \Gamma P_e P(\bm{v},T) ~\mathrm{d}\bm{v}.
\label{eq:He}
\end{equation}
in temperature variation per unit time. The crossing between $H_e$ and $G$ gives the low temperature equilibrium when there is no rf-heating. Depending on the chosen laser parameters, it is between 0.5~mK (the Doppler laser limit) and 5~mK. 

\begin{figure}[hbt]
   \includegraphics[width=0.49\textwidth]{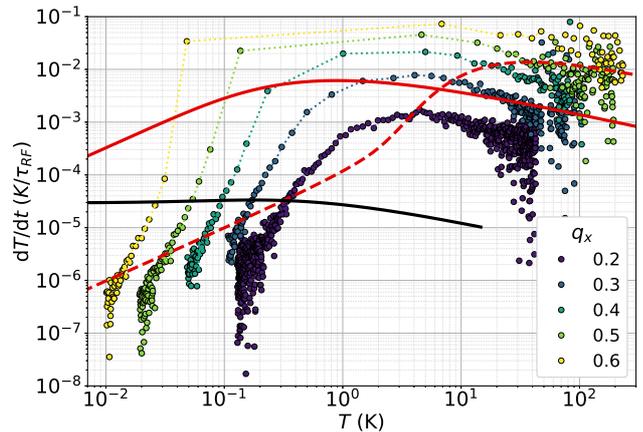}
  \caption{Log-log representation of $\mathrm{d}T/\mathrm{d}t$ {\it vs} temperature in a time-scale in units of the rf period $\tau_{RF}$. Coloured symbols : free evolution $H$ of an ensemble of 1024~ions in an rf trapping potential characterised by  $q_x=0.2, 0.3, 0.4, 0.5, 0.6$ (see the colour chart, the dotted lines are guide to the eyes). Red lines :  cooling rates $G$ as defined by Eq.~(\ref{eq:G}) for different laser detuning : solid line : $\delta_L=-\Gamma$, dashed line : $\delta_L=-10 \Gamma$. Black line : spontaneous emission induced heating $H_e$ as defined by Eq.~(\ref{eq:He}) for $\delta_L=-\Gamma$.}
  \label{fig:heating_rate}
\end{figure}

\subsection{Competition between rf heating and laser cooling}
The comparison between the rf-heating $H$ and laser-cooling rate $G$ (see Fig.~\ref{fig:heating_rate}) shows that, except for the smallest value of the Mathieu parameter $q_x$, these two curves cross in the low temperature regime ($T<1$~K) as well as in the high temperature regime  ($T\simeq 100$~K), the rf-heating being dominant between these two borders. We can then define two possible equilibrium temperatures corresponding to $G=H$, responsible for the bistable behavior of these systems. For initial conditions such that $H>G$, the ensemble converges towards the hotter equilibrium, on the contrary, if $G>H$ the cloud  converges towards the colder equilibrium. For the low temperature range the cold temperature equilibrium is then defined by $G=H_e$. For the largest $q_x$ values, once the low temperature unstable equilibrium $G=H$ is crossed, $H$ reaches its maximum value in a time of the order of magnitude of the averaging window (100 $\mu$s), which is small with respect to the time elapsed since the beginning of the evolution of the ion cloud at low temperature, hence this phenomena usually quoted as an "explosive onset". For increasing $q_x$, the limit temperature for which rf-heating dominates laser-cooling decreases, from 1~K for $q_x=0.3$ to $40(\pm10)$~mK for $q_x=0.6$. This illustrates how the trapping parameter $q_x$ controls the switching conditions between a low and a high temperature regime.

For the chosen ion number $N=1024$ and laser parameters, laser-cooling always overcome rf-heating for $q_x=0.2$ except maybe for temperature of the order of 100~K where the numerical calculations show large fluctuations and are not very relevant. When maximum, the rf-heating rate is two to three orders of magnitude larger than spontaneous emission heating rate. It is also the case for larger $q_x$ values when rf-heating overcome laser-cooling. This hierarchy tells that spontaneous emission heating can be neglected when the competition between the two other effects is considered and confirms the role of $q_x$ parameter as the controlling figure for rf-heating. 

The effect of the number of ions on the general trends detailed above is analysed on Fig.~\ref{fig:heatingN} where the rf-heating rate is plotted for $N=256, 512, 1024$ ions and $q_x=0.6$. It shows that the behaviour identified in Fig.~\ref{fig:heating_rate} is very general for spherical shape of ion ensemble as it does not depend significantly with the ion number.

Furthermore, the comparison between $H$ and $G$ for this high $q_x$ regime, as shown in Fig.~\ref{fig:heating_rate} and \ref{fig:heatingN}, shows that in the intermediate temperature regime where $H>G$, the rf-heating rate is typically one order of magnitude larger than the laser-cooling rate. This justifies in which conditions the impact of laser-cooling on the cloud dynamics can be neglected in numerical simulations as well as in experiments.
\begin{figure}[!htb]
 \includegraphics[width=0.47\textwidth]{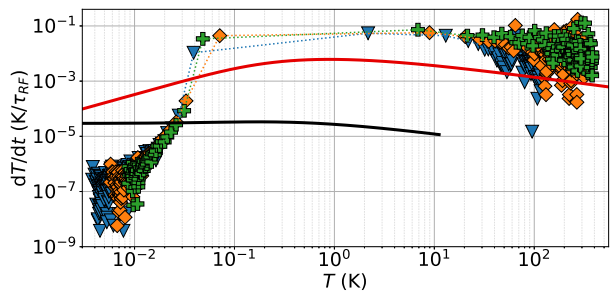}
  \caption{Log-log representation of $\mathrm{d}T/\mathrm{d}t$ {\it vs} temperature in a time-scale in units of the rf period $\tau_{RF}$. Colored symbols : free evolution of $N$~ions and a rf trapping potential characterised by $q_x=0.6$. Blue triangles : $N=256$, Orange diamonds : $N= 512$, Green crosses : $N=1024$ (the dotted lines are guide to the eyes). Red solid lines :  cooling rate $G$ as defined by Eq.~(\ref{eq:G}) for detuning $\delta_L=-\Gamma$. Black line : spontaneous emission induced heating $H_e$ as defined by Eq.~(\ref{eq:He}) for $\delta_L=-\Gamma$.}
  \label{fig:heatingN}
\end{figure}

\subsection{Effect of the density}
For this article, $q_x$ and $N$ were varied by factor, respectively, 3 and 4 and the comparison of Fig.~\ref{fig:heating_rate} and \ref{fig:heatingN} shows that increasing $q_x$ has a stronger impact on the rf heating rate than increasing $N$. For $N=1024$~ions, the radial size of the ion ensemble decreases from $R=190~\mu$m for $q_x=0.6$ to $R=88~\mu$m for $q_x=0.2$. For a given trapping condition $q_x$, this radial size scales like $N^{1/3}$ because we chose to keep the cloud spherical. Studying the dependency of the maximum rf electric field reached by the ions as a function of $q_x$ and $N$ allows to show that this field is not the controlling parameter for the heating rate.

Starting from Eq.~(\ref{eq:rad_pot}), the electric field amplitude $||\bm{E}(r)||$ controlling the motion of an ion at distance $r$ to the trap's central axis, scales like $q_x r$. In the crystal and liquid phase ($T\le 2$~K for the considered densities) \cite{schiffer02}, the density can be considered as homogenous out of a boundary layer of the order of  few $\mu$m \cite{prasad79, champenois09}. The mean density $n_c$ scaling  with $q_x^2$ \cite{prasad79, hornekaer02} and the ion ensemble being spherical, $NR^{-3} \propto q_x^2$. Thus, the maximum rf field amplitude for an ion cloud scales like
\begin{align}
||\bm{E}(R)|| &\propto q_x^{1/3} N^{1/3} \label{eq:Er0}
\end{align}
Our results demonstrate how $q_x$ has a much more significant effect than $N$ on the rf heating rate. Nevertheless, Eq.~(\ref{eq:Er0}) emphasizes how $q_x$ and $N$ play a similar role in the maximum electric field reached by the ions. Thus, it is clear that the parameter controlling the rf heating rate is not   $||\bm{E}(R)||$.  To the contrary, considering that rf heating is induced by ion-ion collisions, the strong impact of $q_x$ could be attributed to the change in the mean density which lead to an ion-ion mean distance scaling like $q_x^{-2/3}$ and is thus reduced by a factor 2.1 by going from $q_x = 0.2$ to $q_x = 0.6$.

\section{Insight into different experimental situations}\label{s:experiments}

In several situations, the  collected ion fluorescence is the only non-destructive signal that can be collected from the ion sample and depending on the laser detuning with the atomic transition, this laser-ion coupling  induces Doppler laser-cooling \cite{metcalfbook}. 

For a constant laser detuning, these systems show a bi-stable behaviour for the temperature, triggered by the Mathieu parameter $q_x$, which catch the strength of the rf-trapping field. If the $q_x$ parameter is kept constant, this strong non-linear behaviour shows that a small perturbation in the temperature can be made sufficient to trigger a switch from the low to the high temperature equilibrium. This perturbation amplification process is at the core of a detection protocol that was proposed to detect giant molecular ions \cite{brevet14,poindron21}. While the temperature increase induced by the energy lost by the projectile $\Delta E$ is too small to be detected by current laser-induced fluorescence-counting techniques, it can bring the trapped ion cloud to a temperature where the rf-heating rate is orders of magnitude larger, triggering the switch to the high temperature state. In this thermal state, the Doppler effect results in the reduction  of  the  laser induced fluorescence rate that is detectable and a saturation parameter $s=2$ and detuning $\delta_L=- \Gamma$  were identified as the best compromise for a high sensitivity of the fluorescence signal to the cloud thermal state \cite{poindron21}. In practice, the energy transfer must bring the temperature above the explosive onset threshold where $G=H$. This temperature threshold decreases with increasing $q_x$ and was found to be 20~mK for $q_x=0.68$, 40~mK for $q_x=0.6$, 100~mK for $q_x=0.5$ and 250~mK for $q_x=0.4$. This analysis justifies the need for a large Mathieu parameter for a very sensitive detector, and the temperature threshold determined according to the method developed in this article are compatible with the previous interaction simulations carried out for $q_x > 0.5$ \cite{poindron21}. Considering the initial temperature of the ensemble negligible compared to the temperature threshold, the smallest $q_x$ value for which a detection is observed can produce a measurement of the temperature increase $\Delta T$ by comparison with numerical data like the one of Fig.~\ref{fig:heating_rate}. As already discussed, the threshold temperature does not depend on the cloud size for typical sizes of hundreds of ions. In a stationary description of the system where we can assume that the ensemble thermalises after the perturbation by the projectile, we can write $\Delta E=N k_B \Delta T$ and by measuring the number of ions in the cloud \cite{kamsap15t}  the energy lost $\Delta E$ can be inferred. This control would turn this giant molecular detector into a device able to measure the stopping power of a strongly correlated non-neutral plasma \cite{zwicknagel99,bussmann06}. In the small Mathieu parameter regime, the conditions for an intruder to be detected in the observed fluorescence light, assuming no rf-heating is amplifying the perturbation, are analysed in \cite{gajewski22}.

Once a detection has been effective, the ion cloud has reached a temperature of the order of 100~K and needs to be cooled down to a temperature lower than the chosen threshold for the next detection, which is lower than 100~mK. For this preparation stage, the bi-stability analysis is also very useful. As shown in Fig.~\ref{fig:heating_rate}, for a very detuned cooling laser ($\delta_L=-10\Gamma$), the laser-cooling $G$ overcomes rf-heating $H$ for large temperature : above 4~K for $q_x=0.2$ and 20~K for $q_x=0.4$. It is then a relevant detuning for cooling ensembles above 100 K but
in order to reach the low temperature equilibrium, decreasing the detuning alone is insufficient and the Mathieu parameter must be also decreased to reach $q_x<0.3$. With such a low Mathieu parameter, there is always a detuning $\delta_L$ such that $G > H$. It is then  possible to bring the ion ensemble to the low equilibrium temperature by gradually changing the laser detuning from a high $|\delta_L|$ to low $|\delta_L|$ value. Then, $q_x$ can be increased up to reach the required detection sensitivity, as long as $G > H$ is verified at the given temperature. 


\section{Conclusion}
This article demonstrates the bistable behaviour of a laser-cooled 3D large ion cloud, trapped in a linear rf-trap. The comparison between different trapping conditions and cloud sizes allows to conclude that the characteristic figure controlling the rf heating rate is the cloud density. This analysis allows to build a protocol to keep cold ion cloud in a high trapping field regime and to evaluate the maximum perturbation that a cloud can absorb without switching to a high temperature state. Conversely, the possibility to trigger the switch from a low to a high temperature regime is analysed in the scope of a detector for giant molecule based on the modification of the cloud fluorescence rate. This detector will take advantage of the amplifying effect  induced by the temperature switch, to be sensitive to the crossing of the ion cloud by a single projectile. 

\section*{Acknowledgments} 
The authors thank David Wilkowski for very stimulating discussions which opened new perspectives for the results presented here. This work is financially supported by CNRS-Innovation (project MegaDalton).
\vfill
%

\end{document}